\documentclass[12pt]{article}
\usepackage{amsfonts, amssymb, amscd}
\usepackage{amsmath}
\usepackage[subnum]{cases}
\usepackage{cite}

\def\l{\left}
\def\r{\right}
\def\bl{\Biggl}
\def\br{\Biggr}
\def\nn{\nonumber}
\def\ql{\textquotedblleft}

\def\1o2{{1\over2}}

\def\a{\alpha}
\def\b{\beta}
\def\ab{{\alpha\beta}}
\def\g{\gamma}
\def\G{\Gamma}
\def\d{\delta}

\def\m{\mu}
\def\n{\nu}
\def\mn{{\mu\nu}}
\def\k{\kappa}
\def\la{\lambda}
\def\La{\Lambda}
\def\p{\phi}
\def\pa{\partial}
\def\s{\sigma}

\usepackage{color}

\begin{document}

\title{Kerr-Schild-Kundt Metrics in Generic Gravity Theories with Modified Horndeski Couplings}
\author{ Metin G\"{u}rses$^{(a)}$\footnote{gurses@fen.bilkent.edu.tr}, Yaghoub Heydarzade $^{(a)}$ \footnote{heydarzade@fen.bilkent.edu.tr}\,\, and \c{C}etin
\c{S}ent\"{u}rk$^{(b)}$\footnote{csenturk@thk.edu.tr} \\
{\small (a) Department of Mathematics, Faculty of Sciences}\\
{\small Bilkent University, 06800 Ankara, Turkey}\\
{\small (b) Department of Aeronautical Engineering}\\
{\small University of Turkish Aeronautical Association, 06790 Ankara, Turkey}
}


\begin{titlepage}
\maketitle
\thispagestyle{empty}

\begin{abstract}
The Kerr-Schild-Kundt (KSK) metrics are known to be one of the universal metrics in general relativity, which means that they solve the vacuum field equations of any gravity theory constructed from the curvature tensor and its higher-order covariant derivatives. There is yet no complete proof that these metrics are universal in the presence of matter fields such as electromagnetic and/or scalar fields. In order to get some insight into what happens when we extend the \ql universality theorem" to the case in which the electromagnetic field is present, as a first step, we study the KSK class of metrics in the context of Modified Horndeski theories with Maxwell's field. We obtain exact solutions of these theories representing the $pp$-waves and AdS-plane waves in arbitrary $D$ dimensions.
\end{abstract}

\end{titlepage}


\setcounter{page}{2}

\section{Introduction}

The Kerr-Schild-Kundt (KSK) metrics belong to a very special type-N metrics in general relativity. Recently, it was shown that they are one of the universal metrics in general relativity solving the vacuum field equations of generic theories of gravitation \cite{ggst,gst1,ghst,gst2,gst3,gst4}. As examples the solutions of the field equations of quadratic, cubic, Born-Infeld, topologically massive gravity,  and $f(Riemann)$ theories of gravitation are given explicitly. In the general case the field equations reduce to $N$-number of Klein-Gordon equations, where $N$ is related to the degree of nonlinearity in the theory.

A generalization of the \ql universality theorem" given in \cite{gst4} to nonvacuum case has not been given yet. However, there are some partial efforts for the case of generalization of the Einstein-Maxwell field equations \cite{pravda1,pravda2,pravda3}. Our main goal in the present work is to extend these works to the KSK metrics. In this direction, as a first step, we shall consider a generic theory of gravity coupled with Horndeski-type \cite{horn} interaction and a modification of it.
We generalize the theorem on the universality of the KSK metrics in generic gravity theories \cite{ghst,gst2,gst4} to a gravity theory whose action is given as the union of the generic gravity action and modified Horndeski action; i.e., in $D$ dimensions,
\begin{eqnarray}\label{actgen}
&&I=\int d^Dx \sqrt{-g}\bl[f(g,R,\nabla R,\ldots,\nabla\nabla\ldots\nabla R,\ldots)\nonumber\\
&&~~~~~~~~~~~~~~~~~~~~~~-\frac{1}{4}F_\mn F^\mn+ \sigma_{1}\, \mathcal{R}_{~~\ab}^\mn F_\mn F^\ab +\sigma_{2}\,R_{~~\ab}^\mn F_\mn F^\ab \br],
\end{eqnarray}
where $f$ is an arbitrary function of the metric tensor ($g_\mn$), the Riemann tensor ($R_{\mn\ab}$), and the covariant derivatives of the Riemann tensor of any order. The term parametrized by $\s_1$ in (\ref{actgen}) represents the Horndeski-type interaction and the term parametrized by $\s_2$ represents the modification of it. The explicit definitions of the tensors $F_\mn$ and $\mathcal{R}_{\mn\ab}$ are given in Section 2, see Eqs. (\ref{F}) and (\ref{cR}). We show that, in the case of the KSK metrics,
the field equations of the generic theory defined by the action (\ref{actgen}) reduce to a system of coupled linear partial differential equations.

The field equations derived from the action (\ref{actgen}) are as follows
\begin{eqnarray}
&&E_{\mn}=\frac{1}{2}(T_\mn+\sigma_{1} \tau_\mn+\sigma_{2}\tilde{\tau}_\mn),\label{eqn001}\\
&&\nabla_\n \mathcal{F}^\mn=0,\label{eqn002}
\end{eqnarray}
where $E_{\mu \nu}$ is the tensor obtained when $f$ in (\ref{actgen}) is varied with respect to $g_{\mu \nu}$, and ${\cal{F}}_{\mu \nu}$ is the tensor when the total Lagrange
function is varied with respect to $F_{\mu \nu}$. The explicit form of the tensors $T_\mn$, $\tau_\mn$, $\tilde{\tau}_\mn$, and ${\cal{F}}_{\mu \nu}$  in (\ref{eqn001}) and
(\ref{eqn002}) are given
in the next section, see Eqs. (\ref{TEM})-(\ref{cF}). In the generic case (with no electromagnetic fields), when the metric is the KSK metric, i.e., of the form $g_{\mu
\nu}=\tilde{g}_{\mu \nu}+2V l_{\mu} l_{\nu}$ (see Section 3), the tensor $E_{\mu \nu}$ on the LHS of (\ref{eqn001}) takes the form (see, e.g., \cite{ghst,gst2,gst4})
\begin{equation}\label{LHS}
E_{\mu \nu}=e g_{\mu \nu}+B l_{\mu }l_{\nu},
\end{equation}
where $e$ is a constant depending on the coupling parameters of the $f$ function of an explicit gravity theory and $B$ is a scalar function given by
\begin{equation}
B\equiv \sum_{n=0}^{N}\,a_{n}(-1)^n \l(\mathcal{O}-2K\r)^{n}\mathcal{O}V. \label{BKS}
\end{equation}
Here, $N$ is related to the number of covariant derivatives of the Riemann tensor that may appear in the $f$ function of a generic gravity theory. In other words, $2N+2$ is
the derivative order of the pure gravity theory defined by the $f$ function in (\ref{actgen}); for example, $N=0$ corresponds to Einstein's gravity (or the
Einstein-Gauss-Bonnet theory), $N=1$ to the quadratic gravity (or more generally $f(Riemann)$ theory), and $N=2$ to the sixth-order gravity theory \cite{gst3}. The constants
$a_n$'s in (\ref{BKS}) are functions of the parameters of the explicit gravity theory at hand, and the linear operator $\mathcal{O}$ appearing in (\ref{BKS}) is defined in
(\ref{rho}) in Section 3.

In section 4, we will show that, when the vector potential is $A_{\mu}=\phi(x) l_{\mu}$, see Eq. (\ref{pot}), the RHS of the equation (\ref{eqn001}) takes the form
\begin{equation}\label{RHS}
RHS\equiv B_{1} l_{\mu } l_{\nu},
\end{equation}
where $B_{1}$ is a scalar function given explicitly in Section 4, see Eqs. (\ref{TEMKS}), (\ref{THKS}), and (\ref{TtildeKS}). Then the field equations (\ref{eqn001}) of the
full theory (\ref{actgen}) reduce to, from (\ref{LHS}) and (\ref{RHS}),
\begin{equation}\label{e=0}
e=0,
\end{equation}
which determines the effective cosmological constant in terms of the parameters of the theory, and
\begin{equation}\label{BB1}
B=B_1,
\end{equation}
which gives higher order linear coupled partial differential equations for $V$ and $\phi$.

To see the simplification introduced by the KSK metrics, let us give the
following two specific examples.

\vspace{0.5cm}
\noindent
{\bf A. Einstein Gravity with Modified-Horndeski Couplings}\\

In this case, the function $f$ appearing in the action (\ref{actgen}) is
\begin{equation}\label{Hilb}
f(g,R)=\frac{1}{2\k^2}(R-2\La),
\end{equation}
which is the Einstein-Hilbert Lagrange function with a cosmological constant. Therefore, from (\ref{LHS})-(\ref{BB1}), we have
\begin{eqnarray}
&&e\equiv\frac{1}{2\k^2}\l[\La-\frac{(D-1)(D-2)}{2}K\r]=0, \\
&&\frac{1}{2\k^2}\mathcal{O}V=\frac{1}{2}\bl\{\psi-4\s_1\bl[\frac{1}{2}\bar{\Box}\psi+\xi^\a\pa_\a \psi-(\xi_\a p^\a)^2+p^\a p^\b\bar{\nabla}_\a\xi_\b
+\frac{1}{2}\l[\xi^2+(D^2-7D+8)K\r]\psi\br]\nn\\
&&~~~~~~~~~~~~~~~~~~+\s_2\bl[\bar{\Box}\psi+2\xi^\a\pa_\a \psi+\frac{1}{2}(\xi_\a p^\a)^2+p^\a p^\b\bar{\nabla}_\a\xi_\b-6K\psi\br]\br\}\label{EKS},
\end{eqnarray}
where we have taken $a_0\equiv\frac{1}{2\k^2}$ and $N=0$ in (\ref{BKS}). In this work, we shall consider this theory  and present the specific solutions to (\ref{EKS}),
together with (\ref{eqn002}).

\vspace{0.5cm}
\noindent
{\bf B. Quadratic Gravity with Modified-Horndeski Couplings}\\

This time the function $f$ in (\ref{actgen}) is
\begin{equation}\label{QLag}
f(g,R)=\frac{1}{2\k^2}(R-2\La_0)+\a R^2+\b R^2_\mn+\g(R^2_{\ab\mn}-4R^2_\mn+R^2),
\end{equation}
which is the combination of the Einstein-Hilbert Lagrangian (with the bare cosmological constant $\La_0$) and the squared terms parameterized by ($\a,\b,\g$). The highest
derivative order of this theory is $N=1$ \cite{gst1}. Then, (\ref{e=0}) determines the cosmological constant in the theory as
\begin{equation}\label{}
e\equiv\frac{\La_0-\La}{2\k^2}-h\La^2=0,
\end{equation}
where we made the definitions
\begin{equation}\label{}
\La\equiv\frac{(D-1)(D-2)}{2}K, ~~~~h\equiv(D\a+\b)\frac{(D-4)}{(D-2)^2}+\g\frac{(D-3)(D-4)}{(D-1)(D-2)},
\end{equation}
and (\ref{BB1}) becomes
\begin{eqnarray}
&&-\b\l[\mathcal{O}-\l(2K+\frac{c}{\b}\r)\r]\mathcal{O}V\nn\\
&&~~~~~~=\frac{1}{2}\bl\{\psi-4\s_1\bl[\frac{1}{2}\bar{\Box}\psi+\xi^\a\pa_\a \psi-(\xi_\a p^\a)^2+p^\a p^\b\bar{\nabla}_\a\xi_\b
+\frac{1}{2}\l[\xi^2+(D^2-7D+8)K\r]\psi\br]\nn\\
&&~~~~~~~~~~~~~~+\s_2\bl[\bar{\Box}\psi+2\xi^\a\pa_\a \psi+\frac{1}{2}(\xi_\a p^\a)^2+p^\a p^\b\bar{\nabla}_\a\xi_\b-6K\psi\br]\br\},\label{QKS}
\end{eqnarray}
where
\begin{equation}\label{}
c\equiv\frac{1}{2\k^2}+\frac{4\La D}{D-2}\a+\frac{4\La}{D-1}\b+\frac{4\La(D-3)(D-4)}{(D-1)(D-2)}\g.
\end{equation}
Note that, in order to obtain the LHS of (\ref{QKS}), one should write (\ref{BKS}) for both the Einstein-Hilbert part (for which $N=0$) and the squared terms part (for which
$N=1$) of (\ref{QLag}), with the redefined $a_n$'s for each. Also observe that, if we set the couplings $\a$, $\b$, and $\g$ to zero, we recover the case A discussed above.

In this work, we will solve the field equations of the gravity theory defined in the case A above for which $f$ is the Hilbert Lagrange function with a cosmological constant given by (\ref{Hilb}), and the matter energy-momentum tensor comes from the modified Horndeski couplings appearing in (\ref{actgen}). For this purpose we structure the paper as follows. In Section 2, we define the modified Horndeski theory and give its field equations. In Section 3, we review the properties of the KSK metrics. In Section 4, we study the KSK metrics in the context of the modified Horndeski theory. In Section 5 and 6,   $pp$-waves and AdS-plane waves in modified Horndeski theory are discussed, respectively. In Section 7, the solutions of some special cases
are given. Finally, in Section 8, we give our concluding remarks. Throughout the paper, we shall use the metric signature $(-,+,+,+,\ldots)$ .

\section{Modified Horndeski Theory}\label{SmHorn}

The action that generalizes the Einstein-Maxwell theory by including the Horndeski's modification in $D$ dimensions is given by
\begin{equation} 
I=\int d^{D}x\sqrt{-g}\,\l[\frac{R-2\La}{2\k^2}-\frac{1}{4}F_\mn F^\mn+\sigma_{1}\, \mathcal{R}_{~~\ab}^\mn F_\mn F^\ab +\sigma_{2}\, R_{~~\ab}^\mn F_{\mu
\nu}F^{\alpha \beta}\r],\label{action}
\end{equation}
where $\k^2$ is the gravitational constant, $\La$ is the cosmological constant, $\sigma_{1}$ and $\sigma_{2}$ are  coupling constants, $R$ is the Ricci scalar, and
\begin{eqnarray}
&&F_\mn\equiv \nabla_\m A_\n-\nabla_\n A_\m,\label{F}\\
&&\mathcal{R}_{~~\ab}^\mn\equiv-\frac{1}{4}\d^{\m\n\la\s}_{\a\b\rho\tau}R_{~~\la\s}^{\rho\tau},\label{cR}
\end{eqnarray}
with $A_\m$ and $R_{~~\la\s}^{\rho\tau}$ being the electromagnetic vector potential and the Riemann tensor, respectively. The generalized Kronecker delta used here is defined
as
\begin{equation}\label{}
  \d^{\a_1\ldots\a_k}_{\b_1\ldots\b_k}=k!\d^{[\a_1}_{\b_1}\ldots\d^{\a_k]}_{\b_k}=k!\d^{\a_1}_{[\b_1}\ldots\d^{\a_k}_{\b_k]}.
\end{equation}
With the definition (\ref{cR}), one can also write the interaction term in (\ref{action}) as
\begin{equation}\label{}
  \mathcal{R}_{~~\ab}^\mn F_\mn F^\ab=-RF^2+4R_\m^{~\n} F_{\n\a}F^{\m\a}-R^\mn_{~~\ab} F_\mn F^\ab,
\end{equation}
where $F^2\equiv F_\mn F^\mn$ and $R_\m^{~\n}$ is the Ricci tensor.

The field equations of the theory by varying the action (\ref{action}) with respect to the independent variables $g^\mn$ and $A_\m$ can be obtained as
\begin{eqnarray}
&&G_\m^{~\n}+\La \d_\m^\n=\k^2(T_\m^{~\n}+\sigma_{1}\,\tau_\m^{~\n}+\sigma_{2}\,\tilde{\tau}_\m^{~\n}),\label{eqn01}\\
&&\nabla_\n \mathcal{F}^\mn=0,\label{eqn02}
\end{eqnarray}
where
\begin{eqnarray}
&&T_\m^{~\n}\equiv F_{\m\a}F^{\n\a}-\frac{1}{4}\d_\m^\n F^2,\label{TEM}\\
&&\tau_\m^{~\n}\equiv\d^{\n\a\b\g}_{\m\rho\s\tau}\nabla_\a F^{\s\tau}\nabla^\rho F_{\b\g}-4\mathcal{R}_{~~\m\a}^{\n\rho} F_{\rho\k} F^{\a\k},\label{TH}\\
&&\tilde{\tau}_{\mu}^{~\n}\equiv\nabla_{\alpha}\, \nabla_{\beta}\,\l(F^{\alpha}\,_{\mu}\,F^{\beta\nu}+F^{\alpha\nu}\,F^{\beta}\,_{\mu} \r), \nonumber\\
&&~~~~~~~+\frac{1}{2}\delta^\nu_{\mu}R_{\alpha\beta\lambda\gamma
}\,F^{\alpha\beta}\,F^{\lambda\gamma}+\frac{3}{2}F^{\alpha\beta}\left({R^\nu}_{\lambda\alpha\beta}\,{F^{\lambda}}_\mu
+R_{\mu\lambda\alpha\beta}\,F^{\lambda\nu}\right), \label{Ttilde}\\
&&\mathcal{F}^\mn \equiv F^\mn-4\sigma_{1}\mathcal{R}_{~~\ab}^\mn F^\ab +4 \sigma_{2}R_{~~\ab}^\mn F^{\alpha \beta}   .   \label{cF}
\end{eqnarray}
Using the Bianchi identity $R_{\mn[\ab;\g]}=0$ and the fact that $\nabla_\g\d^{\m\n\la\s}_{\a\b\rho\tau}=0$, one can rewrite (\ref{eqn02}) as
\begin{equation}\label{}
 \nabla_\n F^\mn-4\sigma_{1} \mathcal{R}_{~~\ab}^\mn \nabla_\n F^\ab+4 \sigma_{2} \nabla_\n\l (R_{~~\ab}^\mn F^{\alpha \beta} \r) =0.\label{eqn03}
\end{equation}

\section{KSK Class of Metrics}


Suppose that the spacetime is endowed with a metric of the \ql generalized" Kerr-Schild form \cite{ks,gg}
\begin{equation}\label{KS}
  g_{\m\n}=\bar{g}_{\m\n}+2Vl_\m l_\n.
\end{equation}
Here, by the word \ql generalized," we mean that the background metric $\bar{g}_{\m\n}$ is a maximally symmetric spacetime; i.e., its curvature tensor has the specific form
\begin{equation}\label{}
  \bar{R}_\ab^\mn=K\d_\ab^\mn,
\end{equation}
with
\begin{equation}\label{}
  K=\frac{\bar{R}}{D(D-1)}=const.
\end{equation}
It is therefore either Minkowski, de Sitter (dS), or anti-de Sitter (AdS) spacetime, depending on whether $K=0$, $K>0$, or $K<0$.
The scalar field (called profile function) $V(x)$ and the vector field $l^\m$ in (\ref{KS}) satisfy the following conditions
\begin{eqnarray}
&&l_\m l^\m=0,~~\nabla_\m l_\n=\frac{1}{2}(l_\m \xi_\n+l_\n \xi_\m),\label{lxi}\\
&&l_\m \xi^\m=0,~~l^\m\pa_\m V=0,\label{lV}
\end{eqnarray}
where $\xi^\m$ is an arbitrary vector field for the time being. From these relations it follows that
\begin{equation}\label{}
  l^\mu\nabla_\m l_\n=0,~~l^\mu\nabla_\n l_\m=0,~~\nabla_\m l^\m=0.
\end{equation}
Kerr-Schild metrics of the form (\ref{KS}) with the properties (\ref{lxi}) and (\ref{lV}) are called the Kerr-Schild-Kundt (KSK) metrics \cite{ggst,gst1,ghst,gst2,gst3,gst4}. All
the properties (\ref{lxi}) and (\ref{lV}), together with the inverse metric
\begin{equation}\label{KSinv}
  g^{\m\n}=\bar{g}^{\m\n}-2Vl^\m l^\n,
\end{equation}
imply that (see, e.g., \cite{gst1})
\begin{eqnarray}
&&\G^\m_{\m\n}=\bar{\G}^\m_{\m\n},~~l_\m\G^\m_{\a\b}=l_\m\bar{\G}^\m_{\a\b},~~l^\a\G^\m_{\a\b}=l^\a\bar{\G}^\m_{\a\b},\label{Gamma}\\
&&\bar{g}^{\a\b}\G^\m_{\a\b}=\bar{g}^{\a\b}\bar{\G}^\m_{\a\b},\\
&&R_{\m\a\n\b}l^\a l^\b=\bar{R}_{\m\a\n\b}l^\a l^\b=-Kl_\m l_\n,\\
&&R_{\m\n}l^\n=\bar{R}_{\m\n}l^\n=(D-1)Kl_\m,\\
&&R=\bar{R}=D(D-1)K,
\end{eqnarray}
and the Einstein tensor is calculated as
\begin{equation}\label{GKS}
  G_\m^{~\n}=-\frac{(D-1)(D-2)}{2}K\d_\m^\n-\rho\, l_\m l^\n,\\
\end{equation}
with
\begin{equation}\label{rho}
  \rho=\l[\bar{\Box}+2\xi^\a\pa_\a +\frac{1}{2}\xi_\a\xi^\a+2(D-2)K\r]V\equiv -\mathcal{O} V,
\end{equation}
where $\bar{\Box}\equiv\bar{\nabla}_\m\bar{\nabla}^\m$ and $\bar{\nabla}_\m$ is the covariant derivative with respect to the background metric $\bar{g}_{\m\n}$.



\vspace{0.5cm}

\noindent


\section{KSK Metrics in Modified Horndeski Theory}

Let us now assume that the spacetime is endowed with a metric of the form (\ref{KS}) and the electromagnetic vector potential is given by
\begin{equation}\label{pot}
  A_\m=\p(x)l_\m,
\end{equation}
where $\p(x)$ is a scalar function and $l_\m$ is the null vector field satisfying (\ref{lxi}) and (\ref{lV}). This immediately yields
\begin{equation}\label{}
  F_\mn=2p_{[\m}l_{\n]},
\end{equation}
where $p_\m\equiv\pa_\m\p$. Now if we further assume that $l_\m p^\m=0$, then $F_\mn$ becomes null, i.e.,
\begin{equation}\label{}
  F^2\equiv F_\mn F^\mn=0,
\end{equation}
and the null vector field $l_\m$ defines a principal null direction of the electromagnetic field, i.e.,
\begin{equation}\label{}
  F_\mn l^\n=0.
\end{equation}
All these mean that the energy-momentum tensor (\ref{TEM}) of the electromagnetic field is of the form
\begin{equation}\label{TEMKS}
  T_\m^{~\n}=\psi l_\m l^\n,
\end{equation}
where $\psi\equiv p_\m p^\m$, and we have a pure radiation field (null dust).

It is now a matter of computation to show that (\ref{cF}) boils down to
\begin{equation}\label{}
  \mathcal{F}^\mn\equiv[1+4\s_1(D-2)(D-3)K+8\s_2K]F^\mn,
\end{equation}
and (\ref{eqn02}) becomes
\begin{equation}\label{delcF}
  -[1+4\s_1(D-2)(D-3)K+8\s_2K][\bar{\Box}\p+\xi^\n p_\n]l^\m=0,
\end{equation}
which, assuming that the coefficient is not zero, yields
\begin{equation}\label{Boxphi}
  \bar{\Box}\p+\xi^\n p_\n=0.
\end{equation}
On the other hand, when the coefficient is zero, the coupling constants are related to each other in the following way
\begin{equation}\label{sK}
  1+4\s_1(D-2)(D-3)K+8\s_2K=0.
\end{equation}
In this case, the dynamics of the scalar field is completely free, and so, any field $\p(x)$ constitutes a source for the Einstein equations (\ref{eqn01}). From now on, we
shall assume that the coefficient is nonzero and $\p(x)$ should satisfy (\ref{Boxphi}).

After a long calculation, one can also calculate (\ref{TH}) and (\ref{Ttilde}) as
\begin{eqnarray}
&&\tau_\m^{~\n}=-4\bl\{\bar{\nabla}_\alpha p_\b \bar{\nabla}^\a p^\b+\frac{1}{2}\xi^\a\pa_\a \psi-(\xi_\a p^\a)^2+\frac{1}{2}\l[\xi^2+(D-2)(D-3)K\r]\psi\br\}l_\m
l^\n,\label{THKS}\\
&&\tilde{\tau}_\m^{~\n}=\bl\{\bar{\Box}\psi+2\xi^\a\pa_\a \psi+\frac{1}{2}(\xi_\a p^\a)^2+p^\a p^\b\bar{\nabla}_\a\xi_\b-6K\psi\br\}l_\m l^\n,\label{TtildeKS}
\end{eqnarray}
where (\ref{Boxphi}) has been used. Now using (\ref{GKS}), (\ref{TEMKS}), (\ref{THKS}), and (\ref{TtildeKS}), we can write the Einstein equations (\ref{eqn01}) as
\begin{eqnarray}
&&\l[\La-\frac{(D-1)(D-2)}{2}K\r]\d_\m^\n-\rho\, l_\m l^\n\nn\\
&&~~~~~~~~=\k^2\bl\{\psi-4\s_1\bl[\bar{\nabla}_\a p_\b \bar{\nabla}^\a p^\b+\frac{1}{2}\xi^\a\pa_\a \psi-(\xi_\a p^\a)^2
+\frac{1}{2}\l[\xi^2+(D-2)(D-3)K\r]\psi\br]\nn\\
&&~~~~~~~~~~~~~~~~~+\s_2\bl[\bar{\Box}\psi+2\xi^\a\pa_\a \psi+\frac{1}{2}(\xi_\a p^\a)^2+p^\a p^\b\bar{\nabla}_\a\xi_\b-6K\psi\br]\br\}l_\m l^\n,\label{EinKS}
\end{eqnarray}
from which we find that
\begin{eqnarray}
&&\La=\frac{(D-1)(D-2)}{2}K,\label{Lambda}\\
&&\nn\\
&&\bar{\Box}V+2\xi^\a\pa_\a V+\l[\frac{1}{2}\xi_\a\xi^\a+2(D-2)K\r]V\nn\\
&&~~~~=-\k^2\bl\{\psi-4\s_1\bl[\bar{\nabla}_\a p_\b \bar{\nabla}^\a p^\b+\frac{1}{2}\xi^\a\pa_\a \psi-(\xi_\a p^\a)^2
+\frac{1}{2}\l[\xi^2+(D-2)(D-3)K\r]\psi\br]\nn\\
&&~~~~~~~~~~~~~~~+\s_2\bl[\bar{\Box}\psi+2\xi^\a\pa_\a \psi+\frac{1}{2}(\xi_\a p^\a)^2+p^\a p^\b\bar{\nabla}_\a\xi_\b-6K\psi\br]\br\}.\label{EinKS1}
\end{eqnarray}
Using the relation
\begin{equation}\label{}
  \bar{\nabla}_a p_\b \bar{\nabla}^\a p^\b=\frac{1}{2}\bar{\Box}\psi+\frac{1}{2}\xi^\a\pa_\a \psi-(D-1)K\psi+p^\a
  p^\b\bar{\nabla}_\a\xi_\b-p^\b\bar{\nabla}_\b(\bar{\Box}\p+\xi^\n\pa_\n\p),
\end{equation}
together with (\ref{Boxphi}), we can equivalently write (\ref{EinKS1}) as
\begin{eqnarray}
&&\bar{\Box}V+2\xi^\a\pa_\a V+\l[\frac{1}{2}\xi_\a\xi^\a+2(D-2)K\r]V\nn\\
&&~~~~=-\k^2\bl\{\psi-4\s_1\bl[\frac{1}{2}\bar{\Box}\psi+\xi^\a\pa_\a \psi-(\xi_\a p^\a)^2+p^\a p^\b\bar{\nabla}_\a\xi_\b
+\frac{1}{2}\l[\xi^2+(D^2-7D+8)K\r]\psi\br]\nn\\
&&~~~~~~~~~~~~~~~~+\s_2\bl[\bar{\Box}\psi+2\xi^\a\pa_\a \psi+\frac{1}{2}(\xi_\a p^\a)^2+p^\a p^\b\bar{\nabla}_\a\xi_\b-6K\psi\br]\br\}.\label{EinKS2}
\end{eqnarray}
Note that when $\xi_\m=0$, $K=0$, and $\s_2=0$, all these expressions recover the flat background ($pp$-wave) case in Horndeski theory \cite{gh}.

\section{$pp$-Waves in Modified Horndeski Theory}

After having the field equations, let us first study $pp$-waves in the modified Horndeski theory. For this purpose we only need to set $\xi_\m=0$ and $K=0$ in the formulation above. Doing this in (\ref{KS}),
(\ref{KSinv}) and (\ref{Lambda}) immediately leads to \cite{ks,gg}
\begin{equation}\label{pp}
  g_{\m\n}=\eta_{\m\n}+2Vl_\m l_\n,
\end{equation}
for the spacetime metric with the flat background metric $\eta_\mn$ (since $\La=0$) and the inverse metric
\begin{equation}\label{ppinv}
  g^{\m\n}=\eta^{\m\n}-2Vl^\m l^\n.
\end{equation}
And doing the same in (\ref{Boxphi}) and (\ref{EinKS2}) produces
\begin{eqnarray}
&&\bar{\Box}\p=0\label{Boxphipp},\\
&&\bar{\Box}V=-\k^2[\psi-(2\s_1-\s_2)\bar{\Box}\psi].\label{EinKSpp}
\end{eqnarray}
Now the vector field $l^\m$ and the scalar fields $\p(x)$ and $V(x)$ satisfy the conditions
\begin{eqnarray}
&&l_\m l^\m=0,~~\nabla_\m l_\n=0,\label{ldi}\\
&&l^\m\pa_\m \p=0,~~l^\m\pa_\m V=0.\label{phiV}
\end{eqnarray}
If we specifically study in the coordinate system $x^\m=(u,v,x^i)$ with $u$ and $v$ being the double null coordinates and $i=1,\ldots,D-2$ in which the null vector $l_\m$ is
taken to be $l_\m=\d^u_\m$, we can easily show that the two conditions in (\ref{phiV}) give
\begin{equation}\label{}
V=V(u,x^i), ~~~~\p=\p(u,x^i),
\end{equation}
the $pp$-wave metric (\ref{pp}) takes the form
\begin{equation}\label{ppuv}
  ds^2=2dudv+2V(u,x^i)du^2+dx_idx^i,
\end{equation}
and the field equations (\ref{Boxphipp}) and (\ref{EinKSpp}) become
\begin{eqnarray}
&&\nabla_\bot^2\p=0\label{delphipp},\\
&&\nabla_\bot^2V=-\k^2[\psi-(2\s_1-\s_2)\nabla_\bot^2\psi],\label{delVpp}
\end{eqnarray}
where $\nabla_\bot^2\equiv\pa_i\pa^i$ and $\psi=\pa_i\p\pa^i\p$. At this point, we can make a further ansatz
\begin{equation}\label{Vanz}
V(u,x^i)=V_0(u,x^i)-\frac{\k^2}{2}\,\p(u,x^i)^2+\k^2(2\s_1-\s_2)\psi(u,x^i),
\end{equation}
with which Eq. (\ref{delVpp}) becomes
\begin{equation}\label{delV0}
\nabla_\bot^2V_0=0.
\end{equation}
Thus any simultaneous solution of (\ref{delphipp}) and (\ref{delV0}) describes a $pp$-wave metric (\ref{ppuv}) with the profile function (\ref{Vanz}) in the modified Horndeski
theory.

\section{AdS-Plane Waves in Modified Horndeski Theory}

In this section, we shall consider AdS-plane waves for which the background metric $\bar{g}_{\m\n}$ is the usual $D$-dimensional AdS spacetime with the curvature
constant
\begin{equation}\label{curv}
  K\equiv-\frac{1}{\ell^2}=-\frac{2|\La|}{(D-1)(D-2)},
\end{equation}
where $\ell$ is the radius of curvature of the spacetime. We shall represent the spacetime by the conformally flat coordinates  $x^\m=(u,v,x^i,z)$ with
$i=1,\ldots,D-3$ and
the background metric\begin{equation}\label{back}
  d\bar{s}^2=\bar{g}_{\m\n}dx^\m dx^\n=\frac{\ell^2}{z^2}(2dudv+dx_idx^i+dz^2),
\end{equation}
where $u$ and $v$ are the double null coordinates. In these coordinates, the boundary of the AdS spacetime lies at $z=0$.

Now if we take the null vector in the full spacetime of the Kerr-Schild form (\ref{KS}) as $l_\m=\d^u_\m$, then using (\ref{KSinv}) along with $l_\m l^\m=0$
we have\begin{equation}\label{}
  l^\m=g^{\m\n}l_\n=\bar{g}^{\m\n}l_\n=\frac{z^2}{\ell^2}\d^\m_v~~\Rightarrow~~l^\a\pa_\a V=\frac{z^2}{\ell^2}\frac{\pa V}{\pa v}=0~~\&~~
  l^\a\pa_\a \p=\frac{z^2}{\ell^2}\frac{\pa \p}{\pa v}=0,
\end{equation}
which represents that  functions $V$ and $\p$ are independent of the coordinate $v$; that is, $V=V(u,x^i,z)$ and $\p=\p(u,x^i,z)$. Therefore, the full spacetime metric
defined by (\ref{KS})
will be
\begin{equation}\label{AdSwave}
  ds^2=\left[\bar{g}_{\m\n}+2V(u,x^i,z)l_\m l_\n\right]dx^\m dx^\n=d\bar{s}^2+2V(u,x^i,z)du^2,
\end{equation}
with the background metric (\ref{back}). It is now straightforward to show that (see also \cite{gst1})
\begin{equation}\label{lxiAdS}
  \nabla_\m l_\n=\bar{\nabla}_\m l_\n=\frac{1}{z}(l_\m\d^z_\n+l_\n\d^z_\m),
\end{equation}
where we used the second property in (\ref{Gamma}) to convert the full covariant derivative $\nabla_\m$ to the background one $\bar{\nabla}_\m$, and $l_\m=\d^u_\m$ with
$\pa_\m l_\n=0$. Comparing (\ref{lxiAdS}) with the defining relation in (\ref{lxi}), we see that
\begin{equation}
\left.\begin{array}{l}
         \displaystyle\xi_\m=\frac{2}{z}\d^z_\m,\\
         \displaystyle\xi^\m=g^{\m\n}\xi_\n=\bar{g}^{\m\n}\xi_\n=\frac{2z}{\ell^2}\d^\m_z,
\end{array} \right\}~~\Rightarrow~~\xi_\m\xi^\m=\frac{4}{\ell^2},
\end{equation}
where we again used (\ref{KSinv}) together with $l_\m \xi^\m=0$. One can also show that
\begin{equation}\label{dxi}
  \bar{\nabla}_\m \xi_\n=\frac{2}{z^2}(\d^z_\m\d^z_\n-\eta_\mn).
\end{equation}
Thus, for the AdS-plane wave ansatz (\ref{AdSwave}), the equations that must be solved are the equation (\ref{Boxphi}), which takes the form
\begin{equation}\label{Boxphiz}
  z^2\hat{\pa}^2\p+(4-D)z\,\pa_z\p=0,
\end{equation}
where $\hat{\pa}^2\equiv\pa_i\pa^i+\pa_z^2$, and the equation (\ref{EinKS2}), which becomes
\begin{eqnarray}
z^2\hat{\pa}^2V+(6-D)z\,\pa_zV+2(3-D)V
&=&-\k^2\bl\{\l[1+\frac{2\s_{1}}{\ell^2}(D^2-7D+8)+\frac{4\s_{2}}{\ell^2}\r](z\hat{\pa}\p)^2\nn\\
&&~~~~~~~~~-\frac{2\s_1-\s_2}{\ell^2}\l[z^2\hat{\pa}^2+(6-D)z\pa_z\r](z\hat{\pa}\p)^2\nn\\
&&~~~~~~~~~+\frac{4(2\s_1+\s_2)}{\ell^2}(z\pa_z\p)^2\br\},\label{Einz}
\end{eqnarray}
where $(\hat{\pa}\p)^2\equiv\pa_i\p\pa^i\p+(\pa_z\p)^2$.

\section{Solutions in Special Cases}
\subsection{Einstein-Maxwell Theory}
As is obvious from the action (\ref{action}), when the non-minimal couplings $\s_1$ and $\s_2$ become zero, we recover the usual Einstein-Maxwell theory. In this case, the
field equations (\ref{Boxphiz}) and (\ref{Einz}) become
\begin{eqnarray}
&&z^2\hat{\pa}^2\p+(4-D)z\,\pa_z\p=0,\label{EMphi}\\
&&z^2\hat{\pa}^2V+(6-D)z\,\pa_zV+2(3-D)V=-\k^2(z\hat{\pa}\p)^2\label{EMV}.
\end{eqnarray}
Let us consider some solutions to these equations.

\subsubsection{Generic Solution in $D=3$}

In $D=3$, the equations (\ref{EMphi}) and (\ref{EMV}) can be solved exactly because $x^\m=(u,v,z)$ and so $V=V(u,z)$ and $\p=\p(u,z)$. Indeed, the equation (\ref{EMphi})
takes the form
\begin{equation}\label{KGAdS3D}
  z^2\pa_z^2\p+z\pa_z\p=0,
\end{equation}
and has the general solution
\begin{equation}\label{p3Dm0}
  \p(u,z)=a_1(u)+a_2(u)\ln z,
\end{equation}
where $a_1(u)$ and $a_2(u)$ are arbitrary functions. Plugging this into (\ref{EMV}), one has
\begin{equation}
z^2\pa_z^2V+3z\pa_zV=-\k^2a_2(u)^2,\label{EinAdS3Dm0}
\end{equation}
which can easily be integrated to give
\begin{equation}
V(u,z)=b_1(u)+b_2(u)z^{-2}-\frac{1}{2}\k^2a_2(u)^2\ln z,\label{sol3D1m0}
\end{equation}
where $b_1(u)$ and $b_2(u)$ are arbitrary functions. Note that the second term $b_2(u)z^{-2}$ can always be absorbed into the AdS part of the metric (\ref{AdSwave}) by a
redefinition of the null coordinate $v$, which means that one can always set $b_2(u)=0$ without loosing any generality. Thus the metric
\begin{equation}\label{}
  ds^2=g_{\m\n}dx^\m dx^\n=\frac{\ell^2}{z^2}(2dudv+dz^2)+2V(u,z)du^2,
\end{equation}
with the profile function given by (\ref{sol3D1m0}), describes an exact plane wave solution propagating in the AdS background spacetime in three-dimensional Einstein-Maxwell
theory.

\subsubsection{Homogeneous AdS-Plane Waves in $D>3$}

In higher dimensions than three, it is not possible to give general solutions to the equations (\ref{EMphi}) and (\ref{EMV}) because now $x^\m=(u,v,x^i,z)$ with
$i=1,\ldots,D-3$ and in general $V$ and $\p$ are functions of $x^i$ also. But we can obtain a special solution if we assume the functions $V$ and $\p$ are homogeneous along
the transverse coordinates, i.e., $V=V(u,z)$ and $\p=\p(u,z)$. If this is the case, the equation (\ref{EMphi}) becomes
\begin{equation}\label{KGAdSD}
  z^2\pa_z^2\p+(4-D)z\pa_z\p=0,
\end{equation}
which has the general solution
\begin{equation}\label{phiD}
  \p(u,z)=a_1(u)+a_2(u)z^{D-3},
\end{equation}
with the arbitrary functions $a_1(u)$ and $a_2(u)$. With this solution, the equation (\ref{EMV}) then becomes
\begin{equation}
z^2\pa_z^2V+(6-D)z\pa_zV+2(3-D)V=-\k^2(D-3)^2a_2(u)^2z^{2(D-3)},\label{EinAdSD}
\end{equation}
which can also be solved exactly and the solution is
\begin{equation}
V(u,z)=b_1(u)z^{D-3}+b_2(u)z^{-2}-\frac{\k^2(D-3)a_2(u)^2}{2(D-2)}\,z^{2(D-3)},\label{solD1}
\end{equation}
where $b_1(u)$ and $b_2(u)$ are arbitrary functions. Notice that this solution is asymptotically well-behaved as $z\rightarrow0$. Therefore, the metric
\begin{equation}\label{}
  ds^2=g_{\m\n}dx^\m dx^\n=\frac{\ell^2}{z^2}(2dudv+dx_idx^i+dz^2)+2V(u,z)du^2,
\end{equation}
with the profile function (\ref{solD1}) describes an exact plane wave, propagating in the $D$-dimensional AdS background, in Einstein-Maxwell theory.

\subsection{Modified Horndeski Theory }

When the Modified Horndeski interactions are present, i.e. $\s_1\neq0$ and $\s_2\neq0$, which is the case in $D>3$, we have to solve the equations (\ref{Boxphiz}) and
(\ref{Einz}). But as we stated before, it is not possible to obtain the general solutions of these equations, so we shall specialize to the homogeneous case in which the
functions $V$ and $\p$ do not depend on the transverse coordinates $x^i$ of the spacetime coordinates $x^\m=(u,v,x^i,z)$, where $i=1,\ldots,D-3$.

\subsubsection{Homogeneous AdS-Plane Waves in $D>3$}

With the assumption that $V=V(u,z)$ and $\p=\p(u,z)$, the field equations (\ref{Boxphiz}) and (\ref{Einz}) become
\begin{eqnarray}
&&z^2\pa_z\p+(4-D)z\,\pa_z\p=0,\label{EMphi1}\\
&&\nn\\
&&z^2\pa_z^2V+(6-D)z\pa_zV+2(3-D)V=-\k^2\bl\{\l[1+\frac{2\s_{1}}{\ell^2}(D^2-7D+12)+\frac{8\s_{2}}{\ell^2}\r](z\pa_z\p)^2\nn\\
&&~~~~~~~~~~~~~~~~~~~~~~~~~~~~~~~~~~~~~~~~~~~~~~~~~~-\frac{2\s_1-\s_2}{\ell^2}\l[z^2\pa_z^2+(6-D)z\pa_z\r](z\pa_z\p)^2\br\}\label{EMV1}.
\end{eqnarray}
The first equation has the solution
\begin{equation}\label{}
  \p(u,z)=a_1(u)+a_2(u)z^{D-3},
\end{equation}
as in the Einstein-Maxwell case, where $a_1(u)$ and $a_2(u)$ are two arbitrary functions. Inserting this solution into the second equation produces
\begin{eqnarray}
&&z^2\pa_z^2V+(6-D)z\pa_zV+2(3-D)V\nn\\
&&=-\k^2(D-3)^2a_2(u)^2\l[1-\frac{2\s_1}{\ell^2}(D+2)(D-3)+\frac{2\s_2}{\ell^2}(D^2-4D+7)\r]z^{2(D-3)},\label{}
\end{eqnarray}
which can easily be solved to give
\begin{eqnarray}
&&V(u,z)=b_1(u)z^{D-3}+b_2(u)z^{-2}\nn\\
&&~~-\frac{\k^2(D-3)a_2(u)^2}{2(D-2)}\l[1-\frac{2\s_1}{\ell^2}(D+2)(D-3)+\frac{2\s_2}{\ell^2}(D^2-4D+7)\r]\,z^{2(D-3)},\label{solD2}
\end{eqnarray}
where $b_1(u)$ and $b_2(u)$ are arbitrary functions. This solution is asymptotically well-behaved as $z\rightarrow0$, and we can
see that it reduces to the solution (\ref{solD1}) in Einstein-Maxwell case when $\s_1=\s_2=0$. Thus we obtained that the metric
\begin{equation}\label{}
  ds^2=g_{\m\n}dx^\m dx^\n=\frac{\ell^2}{z^2}(2dudv+dx_idx^i+dz^2)+2V(u,z)du^2,
\end{equation}
with the profile function (\ref{solD2}) describes an exact AdS-plane wave solution in the modified Horndeski theory. The AdS-plane wave solution of Horndeski theory
can be obtained by easily setting $\s_2=0$ in (\ref{solD2}).

\section{Conclusion}

 In this work, we studied the field equations of the generic gravity with the electromagnetic field. The theories we considered are the several versions of the generic gravity
 with Horndeski type of couplings. We have reduced the field equations for the case of the Kerr-Schild-Kund (KSK) class of metrics and, giving these field equations explicitly for some special cases, we presented exact solutions representing the $pp$-waves and AdS-plane waves in such theories.


\section*{Acknowledgements}

This work is partially supported by the Scientific and Technological Research Council of Turkey (TUBITAK).

\end{document}